\documentclass{edm_article}

\usepackage[T1]{fontenc}
\usepackage{graphicx}
\usepackage{amsmath}
\usepackage[hidelinks]{hyperref}
\usepackage{enumitem}
\usepackage{booktabs}
\usepackage{multirow}
\usepackage{algorithm}
\usepackage{algorithmic}

\begin{document}

\title{From Tool to Teammate: LLM Coding Agents as Collaborative Partners for Behavioral Labeling in Educational Dialogue Analysis}

\numberofauthors{6}
\author{
\alignauthor Eason Chen\\
       \affaddr{Carnegie Mellon University}\\
       \email{easonc@andrew.cmu.edu}
\alignauthor Isabel Wang\\
       \affaddr{Carnegie Mellon University}
\alignauthor Nina Yuan\\
       \affaddr{Carnegie Mellon University}
\and
\alignauthor Sophia Judicke\\
       \affaddr{Carnegie Mellon University}
\alignauthor Kayla Beigh\\
       \affaddr{Carnegie Mellon University}
\alignauthor Xinyi Tang\\
       \affaddr{Carnegie Mellon University}
}

\maketitle

\begin{abstract}
Behavioral analysis of tutoring dialogues is essential for understanding student learning, yet manual coding remains a bottleneck. We present a methodology where LLM coding agents autonomously improve the prompts used by LLM classifiers to label educational dialogues. In each iteration, a coding agent runs the classifier against human-labeled validation data, analyzes disagreements, and proposes theory-grounded prompt modifications for researcher review. Applying this approach to 659 AI tutoring sessions across four experiments with three agents and three classifiers, 4-fold cross-validation on held-out data confirmed genuine improvement: the best agent achieved test $\kappa=0.78$ (SD$=0.08$), matching human inter-rater reliability ($\kappa=0.78$), at a cost of approximately \$5--8 per agent. While development-set performance reached $\kappa=0.91$--$0.93$, the cross-validated results represent our primary generalization claim. The iterative process also surfaced an undocumented labeling pattern: human coders consistently treated expressions of confusion as engagement rather than disengagement. Continued iteration beyond the optimum led to regression, underscoring the need for held-out validation. We release all prompts, iteration logs, and data.
\end{abstract}


\section{Introduction}

Understanding student behavior in tutoring dialogues is fundamental to improving educational technology. Detecting patterns such as help-seeking, confusion, and disengagement enables researchers to evaluate tutoring effectiveness and design adaptive interventions~\cite{baker2004detecting,aleven2000limitations}. Yet behavioral labeling remains a costly bottleneck: trained coders must manually analyze hundreds of sessions, a process requiring weeks of effort and limiting the scale of educational dialogue research.

Large language models (LLMs) offer a promising solution, achieving strong performance on qualitative coding tasks~\cite{dai2024assessing,dai2023llm}. Recent work has increasingly adopted LLM-assisted and hybrid annotation pipelines for educational data~\cite{dai2024assessing,tran2024analyzing}, reducing but not eliminating the need for human expertise in prompt development. However, developing effective prompts still relies heavily on researcher effort: researchers iterate through trial-and-error, analyze errors, and refine instructions over days or weeks. This bottleneck limits the scalability and reproducibility of LLM-assisted educational research.

We present an alternative approach where LLM coding agents serve not as tools to be configured, but as \textit{research teammates} that autonomously iterate on labeling prompts through systematic error analysis. Applying this methodology to behavioral labeling in AI tutoring dialogues, we address practical questions: How many iterations are needed? How much does it cost? Do different agents produce comparable results? And critically, when should iteration stop?

\textbf{Experiment 1} explored this collaborative workflow. A researcher worked with \textbf{Claude Code} over 21 iterations, improving overall from $\kappa=0.77$ to $\kappa=0.78$, with Follow-up Type improving from $\kappa=0.59$ to $\kappa=0.68$, at a total cost of \$22. Beyond the quantitative improvement, this process yielded a pedagogical insight: the agent's systematic error analysis revealed that human coders consistently labeled expressions of confusion (``I don't know,'' ``I'm stuck'') as engagement rather than disengagement, a pattern not explicitly documented in our original codebook. This raised our core \textbf{R}esearch \textbf{Q}uestion: \textbf{Can LLM agents reliably improve prompts for educational content labeling tasks?}

\textbf{Experiment 2} validated the methodology's generalizability. We assigned the same labeling task to three independent coding agents (\textbf{OpenAI Codex}, \textbf{Claude Code}, and \textbf{Gemini}), each starting from scratch with a minimal baseline prompt. Starting from baseline $\kappa \approx 0.70$, after approximately 10 iterations ($\sim$\$5 per agent), all three agents improved prompts substantially. While development-set performance reached $\kappa=0.91$--$0.93$, our primary evaluation uses held-out cross-validation: 4-fold cross-validation confirmed genuine improvement, with the best agent achieving held-out test $\kappa=0.78$ (SD$=0.08$), matching human inter-rater reliability ($\kappa=0.78$). Importantly, agents independently converged on similar disambiguation rules, suggesting the methodology identifies genuine patterns rather than memorizing specific examples.

We also investigated continued iteration beyond the optimal point. When we extended \textbf{Codex} from v7 ($\kappa=0.93$) through v10, performance regressed to $\kappa=0.91$, demonstrating that more iterations do not always yield improvements.

\textbf{Experiments 3--4} examined whether performance differences stemmed from a ``same-family advantage,'' where agents perform better when the classifier shares their architecture. When we switched to \textbf{Claude Opus 4.5} as classifier (Experiment 3), \textbf{Claude Code} achieved $\kappa=0.91$. When we further tested \textbf{Gemini 3 Pro} as classifier (Experiment 4), \textbf{Claude Code} achieved its best performance ($\kappa=0.89$), while \textbf{Gemini Agent} reached $\kappa=0.92$ with its same-family \textbf{Gemini} classifier. A key finding emerged: same-family advantages manifest primarily as \textit{faster convergence} rather than higher performance ceilings, with \textbf{GPT} classifier achieving the highest final performance across agents.

Our contributions span methodology, empirical findings, and transparency. We introduce a collaborative workflow where LLM coding agents autonomously execute develop-test-analyze cycles while researchers provide theoretical oversight. We provide empirical evidence that iterative refinement enables substantial performance gains, with \textbf{Codex} and \textbf{Gemini} achieving the highest performance ($\kappa=0.93$). Through systematic comparison of three classifiers (\textbf{GPT}, \textbf{Claude}, \textbf{Gemini}), we reveal that classifier choice affects both performance ceiling and convergence speed, with same-family classifiers enabling faster iteration while \textbf{GPT} classifier achieved the highest final performance. Our analysis of iteration dynamics demonstrates that continued iteration can lead to regression, not just diminishing returns, underscoring the importance of using separate validation sets to select the best-performing iteration, as in standard machine learning practice. Along with the paper, we released documentation of iteration logs, costs, and failures to enable replication. Finally, we show how systematic disagreement analysis can surface implicit labeling criteria that human coders apply but have not explicitly documented.

\section{Related Work}

\textbf{Educational Dialogue Analysis.} Dialogue-based tutoring systems have a rich research tradition, from early systems like CIRCSIM-Tutor and AutoTutor to modern LLM-powered chatbots. Analyzing student behavior in these dialogues reveals learning processes invisible in outcome measures alone. Baker et al.~\cite{baker2004detecting} pioneered automated detection of ``gaming the system,'' while Aleven et al.~\cite{aleven2000limitations} distinguished productive help-seeking from answer-seeking behavior. Recent work has extended behavioral detection to confusion, disengagement, and self-regulated learning patterns~\cite{zhang2024llmsrl}. However, behavioral labeling at scale remains labor-intensive: most studies rely on small, manually-coded samples that limit generalizability. Our work addresses this bottleneck by enabling scalable, reliable behavioral coding through LLM agent collaboration.

\textbf{LLMs for Qualitative Coding.} LLMs have shown promise for qualitative analysis in education~\cite{dai2024assessing,dai2023llm}, including classroom discussion assessment~\cite{tran2024analyzing} and self-regulated learning detection~\cite{zhang2024llmsrl}. Recent work has compared LLM and human coding performance~\cite{mcclure2024deductive,ganesh2024prompting}, finding that task-specific finetuning often outperforms in-context learning for nuanced educational coding tasks. However, most studies lack documentation of iterative prompt development. Stamper et al.~\cite{stamper2024enhancing} advocate grounding LLM-based analysis in theoretical frameworks; our methodology embeds help-seeking theory~\cite{aleven2000limitations} throughout the iteration process.

\textbf{Automated Prompt Optimization.} Several methods automate prompt improvement through algorithmic search. APE~\cite{zhou2023large} generates and selects instructions via LLM proposals, DSPy~\cite{khattab2023dspy} compiles declarative programs into optimized prompts, and OPRO~\cite{yang2023large} uses LLMs as optimizers guided by past prompt-score pairs. Our approach shares the goal of automated prompt improvement with these methods. However, they typically optimize for a fixed metric through algorithmic search, whereas our approach leverages coding agents that perform qualitative error analysis, propose theory-grounded modifications, and document their reasoning, enabling researcher oversight and pedagogical validation at each step.

\textbf{Coding Agents as Teammates.} Recent coding agents (\textbf{Codex}, \textbf{Claude Code}) can execute complex programming tasks~\cite{chen2021evaluating}. Concurrent work has explored multi-agent systems that automatically optimize grading guidelines through self-reflection~\cite{chu2025gradeopt}, and LM-based approaches that use model judgments as reward functions for instruction optimization~\cite{heyueya2024evaluating}. We extend this by treating agents as \textit{research teammates} that analyze errors, propose hypotheses, and implement fixes, aligning with Ros\'{e} et al.'s~\cite{rose2019explanatory} call for human-AI collaboration that produces explanatory models.

\textbf{Reliability Standards.} Cohen's Kappa above 0.80 indicates ``almost perfect'' agreement~\cite{landis1977measurement}, yet McDonald et al.~\cite{mcdonald2019reliability} found reliability reporting is inconsistent and missing in many research. Our cross-agent validation offers a new approach: LLM agents as IRR partners for rapid, low-cost iteration.

\section{Context and Data Background}

\subsection{Educational Setting}

We used chatbot interaction data from an undergraduate discrete mathematics course ($\sim$120 students) at a large research university. The course deployed an LLM-powered tutoring system with a ``Socratic guardrail'': the chatbot used hints and guiding questions rather than direct answers. Students accessed the chatbot voluntarily through a web interface while working on homework.

This dataset is particularly valuable for studying behavioral labeling because it captures the tension between students' desire for direct answers and the system's pedagogical constraints. Prior research has raised concerns that conversational LLM support can reduce student-generated reasoning when learners rely on the system for answers rather than engaging in productive struggle~\cite{zhai2024effects}. Our chatbot was designed to redirect answer-seeking behavior toward guided reasoning, creating natural variation in student responses: some students engaged with the scaffolding, while others attempted various strategies to extract solutions. This behavioral diversity makes the labeling task both challenging and educationally meaningful, as correctly identifying answer-seeking versus genuine help-seeking has direct implications for understanding learning processes.

\subsection{Dataset Composition}

Our dataset comprises 659 chatbot sessions collected over two semesters (Fall 2024 and Spring 2025). Sessions ranged from brief clarification requests (1--2 exchanges) to extended problem-solving dialogues (20+ exchanges). The median session length was 6 exchanges, with an interquartile range of 3--12 exchanges. Topics spanned the full curriculum, including logic and propositional reasoning (28\%), proof techniques (35\%), set theory (22\%), and combinatorics (15\%).

For the validation set, we randomly sampled 80 sessions stratified by length (short: 1--4 exchanges, medium: 5--10 exchanges, long: 11+ exchanges) and topic area. Two trained research assistants (RAs), both undergraduate students with tutoring experience, served as coders. Before formal labeling, the RAs and a PhD researcher held two calibration meetings to align on codebook interpretation, practice in sample sessions, and discuss edge cases until reaching a consistent understanding. Only after this alignment process did the RAs independently label the 80 validation sessions. The first author, who is a PhD student researcher, then reviewed all disagreements and established consensus labels through case-by-case discussion with the RAs, documenting the reasoning behind each adjudication.

\subsection{Labeling Framework}

Drawing on help-seeking theory~\cite{aleven2000limitations} and our analysis of common interaction patterns, we defined three labeling dimensions:

\textbf{Student Intent} captures the student's primary goal:
\begin{itemize}[leftmargin=*]
    \item \textbf{Answer-Seeking (AS)}: The student primarily wants the answer with limited interest in understanding. Indicators include direct requests for answers and attempts to extract solutions through indirect questioning.
    \item \textbf{Help-Seeking (HL)}: The student genuinely seeks to understand. They engage with hints, ask clarifying questions, and show willingness to work through problems.
    \item \textbf{Other (OT)}: Off-topic conversations, technical questions, or ambiguous interactions.
\end{itemize}

\textbf{Topic Type} categorizes the mathematical content:
\begin{itemize}[leftmargin=*]
    \item \textbf{Conceptual (C)}: Questions about definitions, theorems, or concepts.
    \item \textbf{Procedural (P)}: Questions about how to solve problems or apply techniques.
\end{itemize}

\textbf{Follow-up Type} characterizes how the student responds to guidance:
\begin{itemize}[leftmargin=*]
    \item \textbf{Engage (E)}: Continues working, asks follow-up questions, or expresses confusion (which indicates continued cognitive engagement).
    \item \textbf{Escalate (EA)}: Explicitly requests more direct help. Example: ``Can you just tell me the answer?''
    \item \textbf{Switch (S)}: Abandons the current problem for an unrelated topic.
\end{itemize}

\begin{figure*}[t]
\Description{Screenshot showing two chatbot conversation panels. Left panel shows a student asking for an explanation of a math concept, with the chatbot providing step-by-step guidance and reflection prompts. Right panel shows a student repeatedly asking for answers despite receiving pedagogical scaffolding, illustrating escalation behavior.}
\centering
\includegraphics[width=0.9\textwidth]{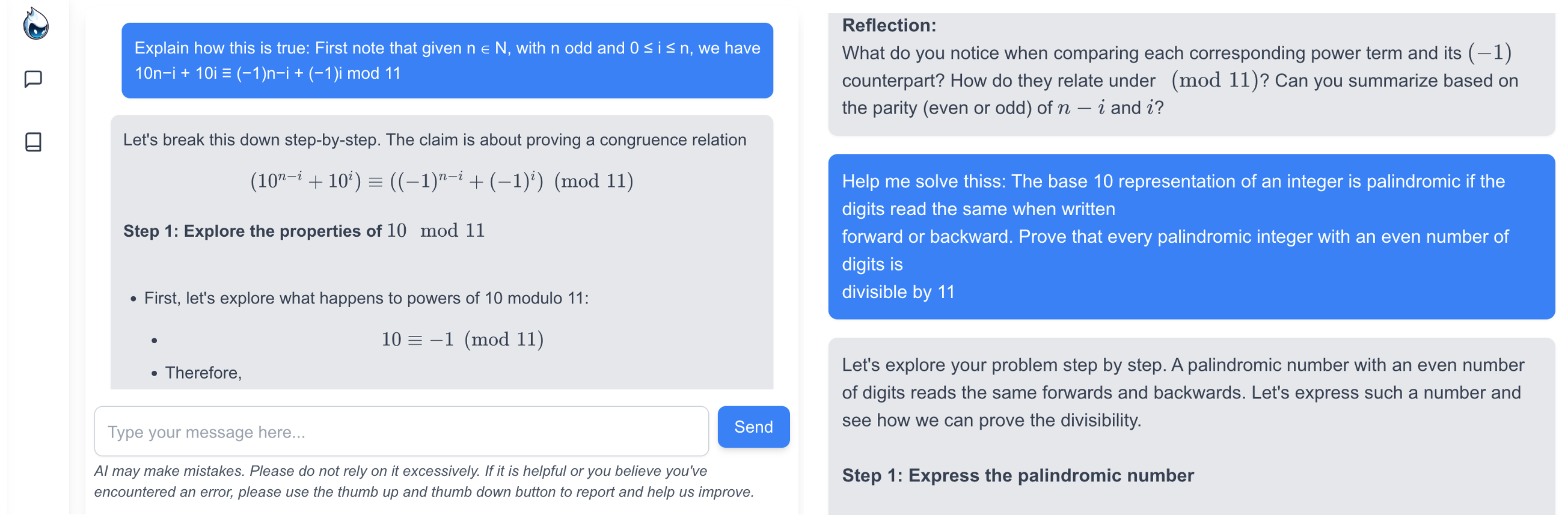}
\caption{Screenshot of the chatbot interface illustrating answer-seeking and escalation behaviors. Left: A student requests an explanation, and the chatbot responds with step-by-step guidance and a reflection prompt. Right: Despite receiving pedagogical scaffolding, the student continues asking for the answer to the same problem, illustrating escalation behavior where students bypass guided reasoning to seek direct solutions.}
\label{fig:chatbot_example}
\end{figure*}

Figure~\ref{fig:chatbot_example} illustrates these behaviors in our chatbot interface. The left panel shows a student engaging with guided reasoning: the chatbot provides step-by-step scaffolding rather than direct answers, and the student works through the problem. The right panel demonstrates escalation: despite receiving pedagogical support, the student continues requesting the answer to a different problem, bypassing the guided reasoning process.

\subsection{Human Coder Reliability}

Our two RAs achieved substantial to almost perfect inter-rater reliability on the 80-session validation set. Student Intent showed $\kappa = 0.78$, Topic Type $\kappa = 0.73$, and Follow-up Type $\kappa = 0.70$, all within the substantial agreement range. When combined across all three dimensions, overall reliability reached $\kappa = 0.89$, placing our coders in Landis and Koch's~\cite{landis1977measurement} ``almost perfect'' agreement category.

These values establish the human baseline that our LLM-based approach aims to match. Notably, Follow-up Type showed the lowest agreement, suggesting this dimension involves the most subjective judgment. This observation foreshadowed the challenges we would encounter during automated labeling. Achieving this level of agreement required extensive discussion during training to align on edge cases, particularly around distinguishing genuine confusion (Engage) from frustrated requests for answers (Escalate).

\section{Experiment 1: Exploratory Development}

\subsection{Method and Workflow}

We used \textbf{GPT-5.2} (gpt-5.2-2025-12-11) as our classifier and \textbf{Claude Code} (powered by \textbf{Claude Opus 4.5}) as our iteration manager. All classifier calls used temperature$=0$ to minimize stochastic variation and ensure reproducible classifications. The agent autonomously executed develop-test-analyze cycles: running the classifier, computing Kappa, identifying disagreement patterns, proposing prompt modifications, and implementing changes. Our role shifted to ``providing oversight'': reviewing changes for pedagogical validity and occasionally vetoing theoretically unsound proposals. Each iteration took 10--15 minutes compute time plus 5--15 minutes human review.

Algorithm~\ref{alg:refinement} formalizes this iterative refinement process.

\begin{algorithm}[t]
\caption{Agent-Driven Iterative Prompt Refinement}
\label{alg:refinement}
\begin{algorithmic}[1]
\REQUIRE Labeled validation set $D$, initial prompt $P_0$, classifier $M$, coding agent $A$
\FOR{$i = 1$ to max\_iterations}
  \STATE Run classifier $M$ with prompt $P_{i-1}$ on $D$
  \STATE Compute $\kappa$ and per-category metrics
  \IF{$\kappa$ plateaus for 2 consecutive iterations}
    \STATE \textbf{break}
  \ENDIF
  \STATE Agent $A$ analyzes disagreements between $M$'s labels and human consensus
  \STATE Agent $A$ proposes prompt modifications with documented reasoning
  \STATE Researcher reviews modifications for theoretical validity
  \STATE $P_i \leftarrow$ approved modifications applied to $P_{i-1}$
\ENDFOR
\RETURN $P^* = \arg\max_{P_i} \kappa(P_i, D)$
\end{algorithmic}
\end{algorithm}

\subsection{Results: 21 Iterations to $\kappa=0.78$}

The iterative process unfolded in three distinct phases.

\subsubsection{Phase 1 (v1--v6): Baseline Calibration}

We initialized the prompt with our codebook definitions (v0), achieving $\kappa=0.77$ out of the box. This surprisingly strong baseline suggests that clear category definitions transfer well to LLM classifiers.

In v1--v2, the agent noticed ``prove this'' requests were misclassified as Answer-Seeking. It added: ``\textit{Requests containing `prove' or `show that' indicate Help-Seeking, as proofs require understanding, not just answers.}'' However, this overcorrected: the classifier then labeled ``prove 2+2=4 for me'' as Help-Seeking when it was clearly Answer-Seeking. Performance dropped to $\kappa=0.71$. We reverted and learned: \textit{keyword-based rules can have unintended consequences}. Versions v3--v6 instead refined Topic Type boundaries, adding: ``\textit{If a question asks both `what' (Procedural) and `why' (Conceptual), classify based on the primary intent indicated by the final sentence.}'' This improved Topic Type $\kappa$ from 0.64 to 0.77.

\subsubsection{Phase 2 (v7--v13): The ``Confusion = Engage'' Discovery}

This phase produced our most significant insight. Follow-up Type showed persistent disagreement around confusion expressions. The original prompt treated ``I don't know'' as potential Escalate (frustration), but human coders consistently labeled these as Engage.

The resolution emerged from examining human reasoning: \textit{admitting confusion is the first step toward learning}. A student saying ``I don't know'' is still engaged, actively communicating their mental state. This contrasts with ``just tell me the answer,'' which represents withdrawal from learning.

This aligns with Kapur's~\cite{kapur2008productive} theory of \textit{productive failure}: struggling with challenges can be more effective than direct instruction. The insight was not in our original codebook; it emerged from systematic analysis of human-AI disagreements.

Version 8 modified the Follow-up Type rules. The original prompt stated: ``\textit{Escalate: Student expresses frustration or requests more help.}'' The agent changed this to: ``\textit{Engage: Student expresses confusion (`I don't know,' `I'm stuck,' `I don't understand') WITHOUT requesting direct answers. Escalate: Student explicitly requests direct answers (`just tell me,' `what's the answer') or expresses frustration WITH a demand.}'' This improved Follow-up $\kappa$ from 0.64 to 0.73. Subsequent iterations refined edge cases: ``I don't know, can you just explain it?'' (Escalate, because of ``just'') versus ``I don't know where to start'' (Engage, confusion without demand).

\subsubsection{Phase 3 (v14--v21): Finalization}

The final phase addressed remaining edge cases: implicit answer-seeking (``Is the answer 42?'' appears as clarification but functions as answer-seeking), rare category stabilization for ``Other,'' and multi-part questions spanning categories.

\subsection{Final Performance}

After 21 iterations, we achieved:
\begin{itemize}[leftmargin=*]
    \item Student Intent: $\kappa = 0.76$ (human: 0.78)
    \item Topic Type: $\kappa = 0.77$ (human: 0.73)
    \item Follow-up Type: $\kappa = 0.73$ (human: 0.70)
    \item Overall: $\kappa = 0.78$ (human: 0.89)
\end{itemize}

The classifier achieved reliability comparable to human coders on individual dimensions. Notably, Topic Type and Follow-up Type showed strong agreement, suggesting that systematic prompt refinement can reduce the subjectivity inherent in these judgments. The gap in overall $\kappa$ (0.78 vs.\ 0.89) reflects accumulated disagreements across dimensions rather than a fundamental limitation.

\subsection{Cost and Reflection}

Total cost was approximately \$22 (API costs plus agent compute time) over 5 hours of researcher oversight. Key success factors included clear category definitions, systematic error analysis, and human oversight catching theoretically unsound proposals. However, questions remained: Did \textbf{Claude Code} find a locally optimal solution? Would other agents discover different solutions? These questions motivated Experiment 2.

\section{Experiment 2: Cross-Agent Validation}

\subsection{Motivation}

Experiment 1's results raised concerns about overfitting: agent-specific artifacts, test set overfitting from 21 iterations on 80 sessions, and potential researcher bias. We designed a replication with multiple independent agents starting from scratch.

\subsection{Method}

We selected three coding agents: \textbf{OpenAI Codex} CLI, \textbf{Gemini} Code Agent, and \textbf{Claude Code}. Each received identical instructions: ``Improve the labeling prompt to maximize Cohen's Kappa. For each iteration: run evaluation, analyze the lowest-$\kappa$ category, identify disagreement patterns, improve the prompt, and document changes.'' Agents had access to the codebook, 80 labeled sessions, and an evaluation script. They worked independently; complete instructions are in supplementary materials. All used \textbf{GPT-5.2} as classifier for comparability. All classifier calls used temperature$=0$ to minimize stochastic variation. While LLM outputs can still vary across runs, our cross-validation across multiple folds provides robustness against single-run artifacts.

\subsection{Results}

\subsubsection{Overall Performance}

\begin{figure}[t]
\Description{Line graph showing Cohen's Kappa values over iterations for agents. \textbf{Codex} reaches 0.93 by iteration 7 then shows slight regression. \textbf{Claude Code} reaches 0.91 by iteration 5. A dashed horizontal line marks the human-human baseline at 0.89.}
\centering
\includegraphics[width=\columnwidth]{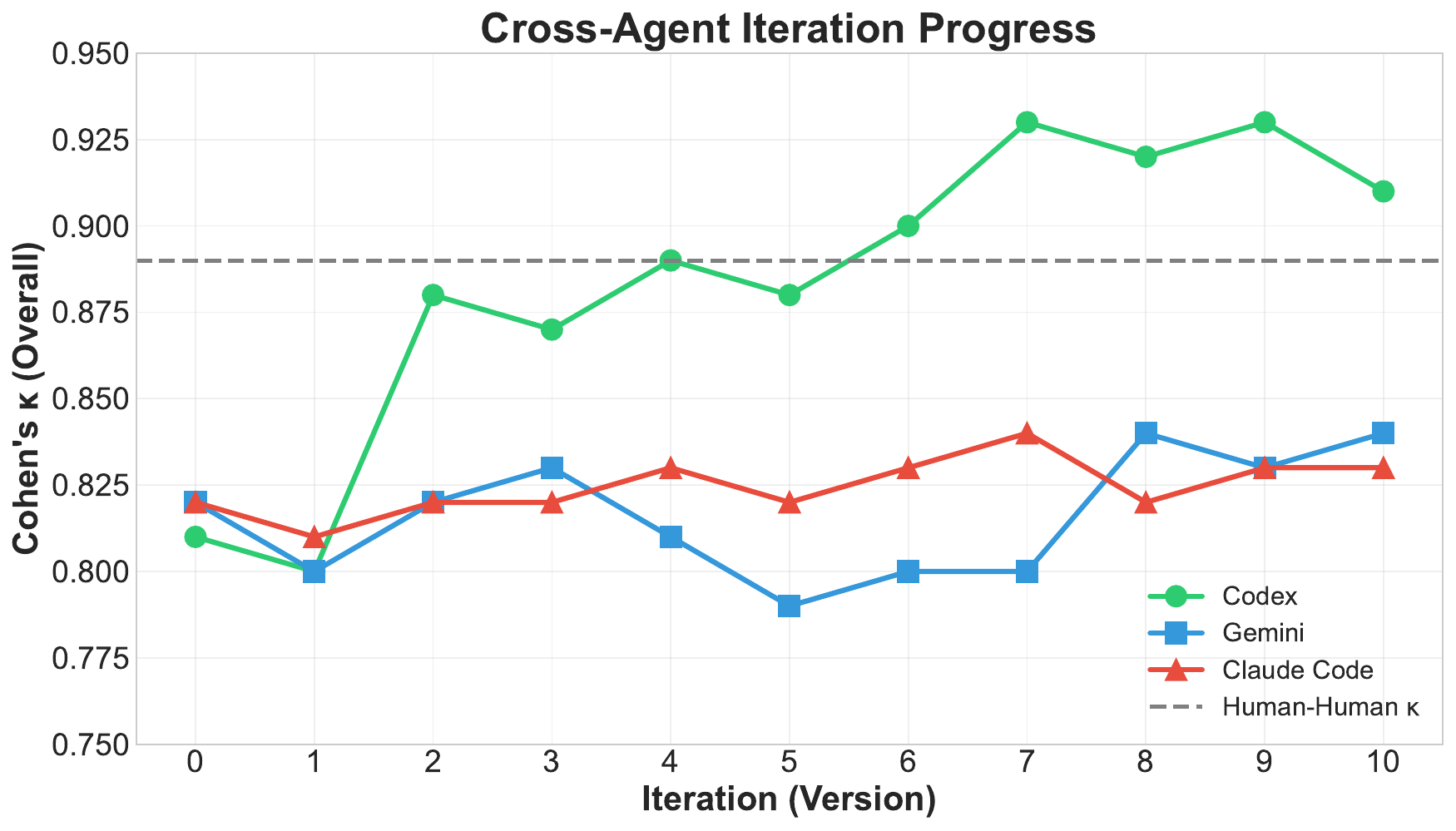}
\caption{Overall $\kappa$ progression across iterations. \textbf{Codex} and \textbf{Gemini} both reached $\kappa=0.93$. \textbf{Claude Code} reached $\kappa=0.91$ at v5. Continued iteration can lead to regression. Dashed line shows human baseline ($\kappa=0.89$).}
\label{fig:progress}
\end{figure}

All three agents achieved strong reliability: \textbf{Codex} and \textbf{Gemini} both reached $\kappa=0.93$ (F1$=0.88$ and $0.91$ respectively), while \textbf{Claude Code} reached $\kappa=0.91$ (F1$=0.79$). Figure~\ref{fig:progress} shows the progression across iterations.

\begin{table}[t]
\centering
\renewcommand{\arraystretch}{1.2}
\caption{Cross-agent validation results. Best performance shown for each agent.}
\label{tab:results}
\begin{tabular}{llllllll} \hline
Agent & Best Ver. & Overall $\kappa$ & F1 & Intent & Topic & Follow-up & Cost \\ \hline
\textbf{Codex} (OpenAI) & v7 & \textbf{0.93} & 0.88 & 0.84 & \textbf{0.85} & 0.83 & \$8 \\
\textbf{Gemini} & v9 & \textbf{0.93} & \textbf{0.91} & \textbf{0.85} & 0.81 & \textbf{0.87} & \$7 \\
\textbf{Claude Code} & v5 & 0.91 & 0.79 & 0.73 & 0.78 & 0.88 & \$7 \\
\midrule
Human-Human & -- & 0.89 & -- & 0.78 & 0.73 & 0.70 & -- \\ \hline
\end{tabular}
\end{table}

Table~\ref{tab:results} presents the final performance. \textbf{Codex} and \textbf{Gemini} tied for highest reliability ($\kappa=0.93$), with \textbf{Gemini} achieving the highest F1 score ($0.91$). \textbf{Claude Code} reached $\kappa=0.91$ (F1$=0.79$). Costs remained modest at \$7--8 per agent, making multi-agent validation economically feasible.

\subsubsection{Per-Dimension Analysis}

\begin{figure}[t]
\Description{Bar chart comparing Cohen's Kappa values across four dimensions (Student Intent, Topic Type, Follow-up Type, Overall) for \textbf{Codex} (OpenAI), \textbf{Gemini} (Google), \textbf{Claude Code} (Anthropic), and Human-Human baseline. \textbf{Codex} and \textbf{Gemini} both achieve 0.93 overall. \textbf{Gemini} leads on Follow-up Type (0.87) and Intent (0.85), while \textbf{Codex} leads on Topic Type (0.85).}
\centering
\includegraphics[width=\columnwidth]{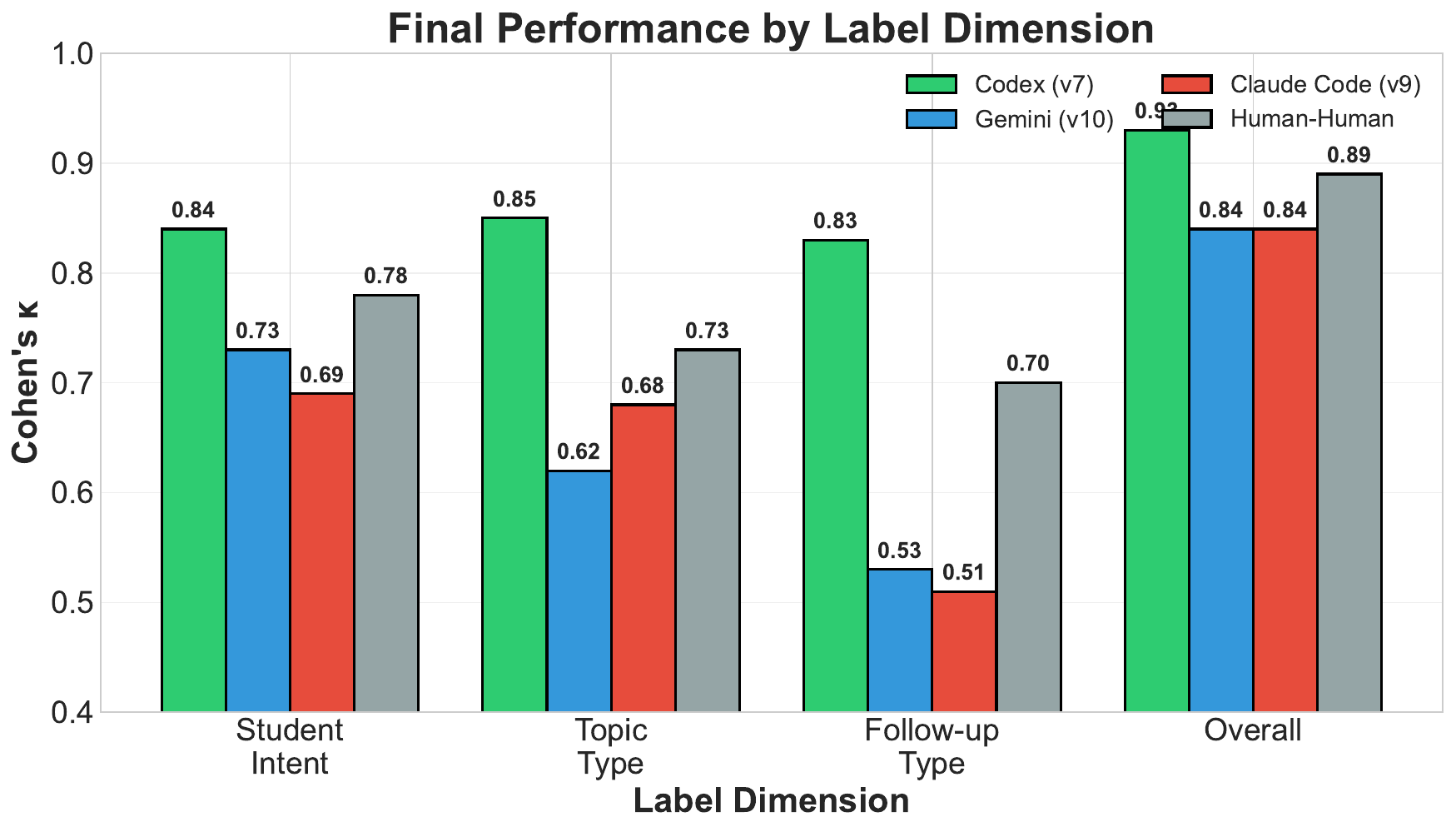}
\caption{Final performance by label dimension. \textbf{Gemini} achieved highest Follow-up Type (0.87), while \textbf{Codex} led on Topic Type (0.85).}
\label{fig:dimensions}
\end{figure}

Figure~\ref{fig:dimensions} shows agents had different strengths. \textbf{Gemini} achieved highest performance on Intent ($\kappa=0.85$) and Follow-up ($\kappa=0.87$), while \textbf{Codex} excelled on Topic Type ($\kappa=0.85$). \textbf{Claude Code} showed lower performance on Intent ($\kappa=0.73$) but comparable Follow-up Type ($\kappa=0.88$).

\subsubsection{The Follow-up Type Bottleneck}

\begin{figure}[t]
\Description{Line graph showing Follow-up Type Kappa progression over iterations. \textbf{Codex} (OpenAI) shows a breakthrough at v7 reaching 0.83. \textbf{Gemini} achieves the highest at v9 with 0.87. \textbf{Claude Code} shows steady improvement reaching 0.88 by v5.}
\centering
\includegraphics[width=\columnwidth]{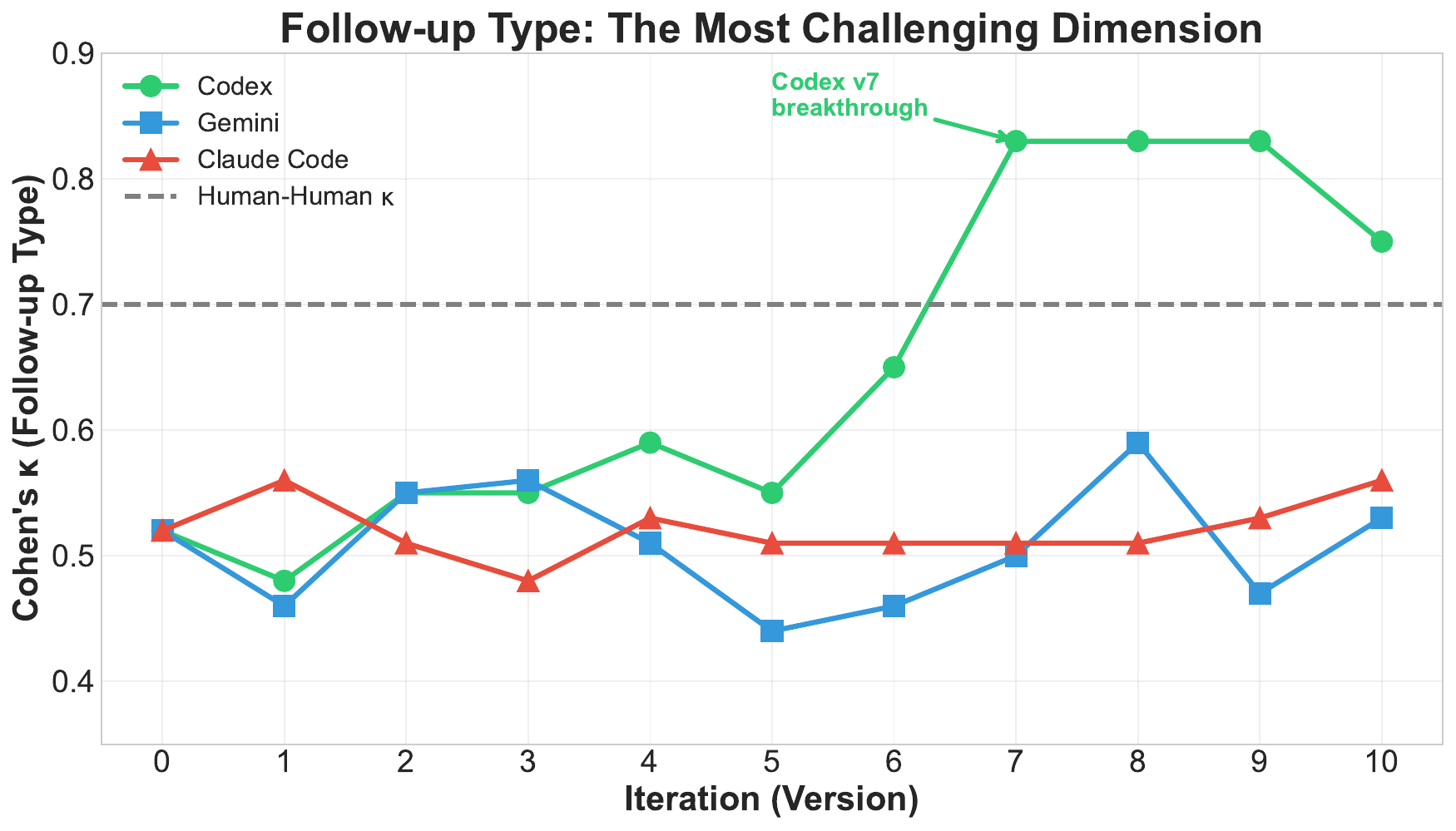}
\caption{Follow-up Type progression. All agents started at $\kappa \approx 0.45$--$0.55$. \textbf{Codex} achieved a breakthrough at v7 (0.83), and \textbf{Gemini} reached $\kappa=0.87$ at v9. \textbf{Claude Code} peaked early at v5 (0.88).}
\label{fig:followup}
\end{figure}

Follow-up Type was initially challenging for all agents, echoing Experiment 1 and human coding patterns. Figure~\ref{fig:followup} shows all agents started with $\kappa \approx 0.45$--$0.55$. \textbf{Codex} achieved a breakthrough at v7, reaching 0.83 by narrowing Switch and defaulting to Engage. \textbf{Gemini} made similar discoveries by v9, achieving the highest Follow-up $\kappa=0.87$. Both agents converged on the key insight: expressions of confusion should be labeled as Engage rather than other categories.

\subsubsection{Convergent Solutions}

Despite working independently, agents converged on similar rules: (1) Switch reserved for truly unrelated topics; (2) Escalate requires explicit answer requests; (3) Engage as default for continued dialogue. These mirror the ``confusion = engagement'' insight from Experiment 1.

This independent convergence is significant: three agents with different architectures, working without access to each other's solutions, discovered similar disambiguation strategies. This suggests the methodology identifies genuine patterns in the data rather than agent-specific artifacts.

\subsection{Iteration Dynamics and Regression}

A notable finding is that continued iteration does not always improve performance. We extended \textbf{Codex} beyond its v7 optimum through v10, observing the following pattern:

\begin{table}[h]
\centering
\renewcommand{\arraystretch}{1.2}
\caption{\textbf{Codex} (OpenAI) iteration dynamics showing regression after v7.}
\label{tab:regression}
\begin{tabular}{llllll} \hline
Version & Overall & Intent & Topic & Follow-up \\ \hline
v7 & \textbf{0.93} & 0.84 & \textbf{0.85} & \textbf{0.83} \\
v8 & 0.92 & 0.81 & 0.80 & 0.83 \\
v9 & 0.93 & 0.84 & 0.84 & 0.83 \\
v10 & 0.91 & 0.81 & 0.85 & 0.75 \\ \hline
\end{tabular}
\end{table}

As shown in Table~\ref{tab:regression}, v8 introduced additional disambiguation rules based on v7's remaining errors, but these changes caused slight regression (0.93 $\to$ 0.92). Version 9 reverted some changes and recovered to 0.93. Version 10 attempted further refinements but caused significant regression in Follow-up Type (0.83 $\to$ 0.75), dropping overall $\kappa$ to 0.91.

This pattern suggests several important lessons:
\begin{itemize}[leftmargin=*]
    \item \textbf{Diminishing returns}: Once high performance is achieved, additional rules may introduce new errors while fixing old ones.
    \item \textbf{Overfitting risk}: Rules developed from error analysis on the validation set may not generalize.
    \item \textbf{Stopping criteria}: Researchers should establish clear stopping criteria rather than iterating indefinitely.
\end{itemize}

\subsection{Cost Analysis}

\begin{table}[t]
\centering
\renewcommand{\arraystretch}{1.2}
\caption{Cost breakdown by agent for Experiment 2.}
\label{tab:costs}
\begin{tabular}{lrr} \hline
Agent & Iterations & Est.\ Cost \\ \hline
\textbf{Codex} (OpenAI) & 11 (v0--v10) & $\sim$\$8 \\
\textbf{Claude Code} & 11 (v0--v10) & $\sim$\$7 \\
\textbf{Gemini} & 11 (v0--v10) & $\sim$\$7 \\
\midrule
\textbf{Total} & 33 & $\sim$\$22 \\ \hline
\end{tabular}
\end{table}

As shown in Table~\ref{tab:costs}, each agent cost approximately \$5--8 for 10+ iterations, totaling around \$22 for all three agents in Experiment 2. The initial exploratory development in Experiment 1 cost approximately \$22 (including \textbf{Claude Code} agent fees and OpenAI API costs over 21 iterations). This modest investment provided three independent validations of the labeling approach, evidence that solutions generalize across agent architectures, and multiple prompt variants that can be compared and combined.

For context, training human coders typically requires 2--4 hours per coder, plus ongoing calibration sessions. Our approach required approximately 5 hours of researcher oversight total across all three agents. The primary researcher time investment was in reviewing agent-proposed changes for pedagogical validity, which we consider essential rather than optional overhead.

\subsection{Cross-Validation}

To assess generalization, we conducted 4-fold cross-validation. We chose 4-fold cross-validation (20 sessions per fold) rather than the conventional 10-fold (8 sessions per fold) because our labeling task requires sufficient category representation within each fold. With only 80 sessions and three multi-class dimensions, 8-session folds risk having zero instances of rare categories (e.g., Switch, Other), making $\kappa$ estimation unreliable. We split the 80 sessions into four folds of 20 sessions each, stratified by intent distribution. For each fold, agents developed prompts using 60 training sessions and evaluated on 20 held-out sessions, with each agent iterating approximately 10 versions per fold (113 total iterations across all agent-fold combinations).

Results confirmed iteration benefits but revealed substantial overfitting. Comparing validation-set performance (Experiment 2's 80 sessions) to cross-validated test performance (held-out 20 sessions per fold) quantifies this gap: \textbf{Codex} achieved validation $\kappa=0.93$ but mean test $\kappa=0.78$; \textbf{Claude} achieved validation $\kappa=0.91$ but mean test $\kappa=0.76$; \textbf{Gemini} achieved validation $\kappa=0.93$ but mean test $\kappa=0.72$. This $\Delta\kappa \approx 0.15$--$0.21$ gap indicates that the high validation-set performance partially reflects overfitting to the 80-session set.

Despite overfitting, \textbf{the methodology demonstrably improves generalization performance}: all agents improved from baseline test $\kappa \approx 0.70$ to test $\kappa = 0.72$--$0.78$, representing genuine gains of $\Delta\kappa = 0.02$--$0.08$ on held-out data. This is the core finding: even after accounting for overfitting, agent-driven iteration produces prompts that generalize better than the baseline. The best cross-validated performance (\textbf{Codex} test $\kappa=0.78$) also matches human inter-rater reliability ($\kappa=0.78$). The variance across folds (SD$=0.08$--$0.12$) reflects sensitivity to which specific sessions appear in training versus test sets.

\begin{table}[h]
\centering
\renewcommand{\arraystretch}{1.2}
\caption{4-fold cross-validation results (test set $\kappa$). Baseline (v0) $\kappa \approx 0.70$ for all agents.}
\label{tab:cv_results}
\begin{tabular}{lccccc} \hline
Agent & Fold 0 & Fold 1 & Fold 2 & Fold 3 & Mean $\pm$ SD \\ \hline
\textbf{Codex} & 0.85 & 0.61 & 0.88 & 0.79 & 0.78 $\pm$ 0.12 \\
\textbf{Claude} & 0.71 & 0.81 & 0.84 & 0.66 & 0.76 $\pm$ 0.08 \\
\textbf{Gemini} & 0.71 & 0.67 & 0.88 & 0.63 & 0.72 $\pm$ 0.11 \\ \hline
\end{tabular}
\end{table}

\section{Experiment 3: Claude Classifier}

\subsection{Motivation}

In Experiment 2, \textbf{Codex} (powered by \textbf{GPT-5.2}) achieved the highest performance among agents when using \textbf{GPT-5.2} as the underlying classifier. This raised an important question: does the agent's superior performance reflect genuine capability differences, or does it stem from a ``same-family advantage'' where agents perform better when the classifier shares their model architecture?

\subsection{Method}

To test this hypothesis, we replicated Experiment 2 using \textbf{Claude Opus 4.5} as the classifier instead of \textbf{GPT-5.2}. The same three agents (\textbf{Codex}, \textbf{Claude Code}, \textbf{Gemini}) independently iterated on the same labeling task for 11 versions (v0--v10), with identical instructions and starting prompts. We verified through transcript analysis that no agent accessed another agent's prompts during iteration.

\subsection{Results}

\begin{table}[t]
\centering
\renewcommand{\arraystretch}{1.2}
\caption{Best performance comparison across Agents and Models.}
\label{tab:classifier_comparison}
\begin{tabular}{llll} \hline
Agent & \textbf{GPT} Classifier & \textbf{Claude} Classifier & $\Delta$ \\ \hline
\textbf{Codex} (OpenAI) & \textbf{0.93} (v7) & \textbf{0.93} (v10) & 0.00 \\
\textbf{Claude Code} & 0.91 (v5) & \textbf{0.91} (v7) & 0.00 \\
\textbf{Gemini} & \textbf{0.93} & 0.85 & $-$0.08 \\ \hline
\end{tabular}
\end{table}

Table~\ref{tab:classifier_comparison} shows the best performance achieved by each agent with each classifier. \textbf{Codex} achieved the same peak ($\kappa=0.93$) with both classifiers, though it required more iterations to reach this level with \textbf{Claude} (v10 vs.\ v7). \textbf{Claude Code} achieved identical peaks with both classifiers ($\kappa=0.91$), reaching this level faster with \textbf{Claude} (v7 vs.\ v5). \textbf{Gemini} performed best with \textbf{GPT} classifier ($\kappa=0.93$), dropping to $0.85$ with \textbf{Claude}.

\begin{figure*}[t]
\Description{Two line graphs comparing iteration progress (v0-v10) for three agents (\textbf{Codex}, \textbf{Claude Code}, \textbf{Gemini}) with different classifiers. Left panel shows \textbf{Claude} classifier results. Right panel shows \textbf{GPT} classifier results with \textbf{Codex} and \textbf{Gemini} both reaching 0.93.}
\centering
\includegraphics[width=0.9\textwidth]{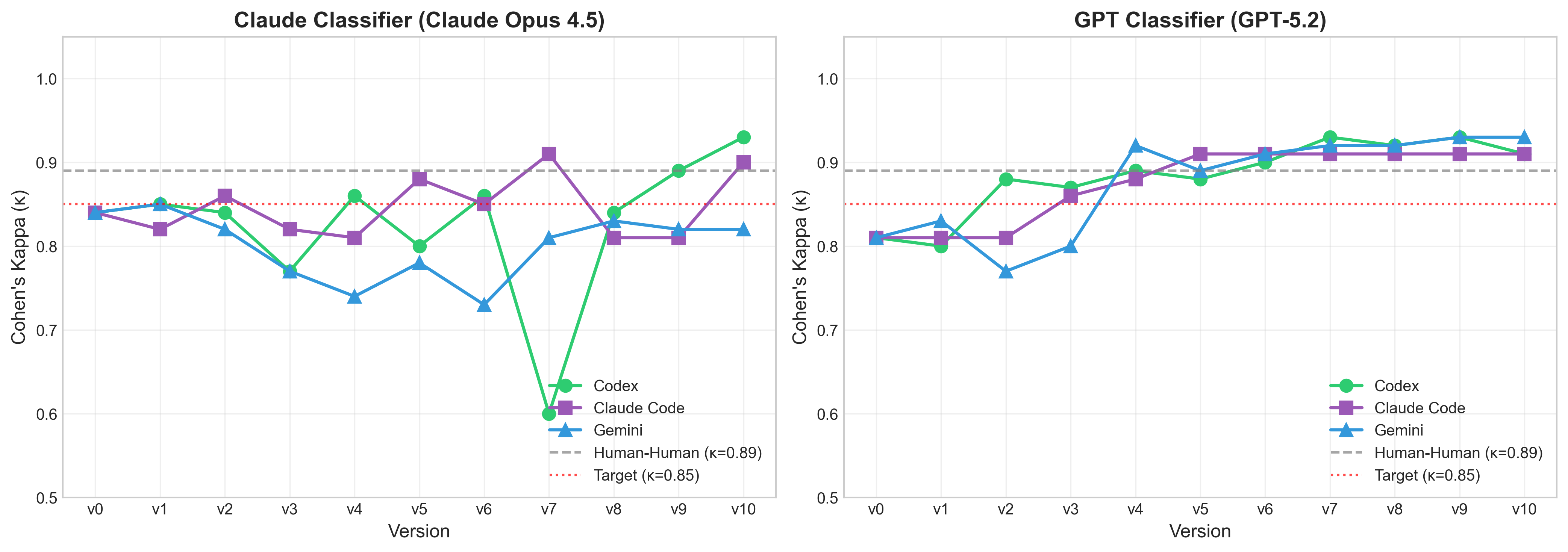}
\caption{Iteration dynamics across classifier models. Left: \textbf{Claude Opus 4.5} classifier. Right: \textbf{GPT-5.2} classifier. \textbf{Codex} achieves $\kappa=0.93$ with both classifiers. \textbf{Gemini} peaks at $0.93$ with \textbf{GPT} but only $0.85$ with \textbf{Claude}.}
\label{fig:classifier_comparison}
\end{figure*}

Figure~\ref{fig:classifier_comparison} shows the iteration dynamics for both classifiers. With \textbf{GPT} classifier, all agents achieved strong performance, with \textbf{Codex} and \textbf{Gemini} both reaching $\kappa=0.93$. With \textbf{Claude} classifier, patterns were more varied, with \textbf{Codex} maintaining $0.93$ while other agents showed lower peaks.

\subsection{Interpretation}

The results partially support the same-family advantage hypothesis:

\textbf{\textbf{Claude Code} shows faster convergence with \textbf{Claude} classifier.} \textbf{Claude Code} achieved identical peak performance with both \textbf{GPT} and \textbf{Claude} classifiers ($\kappa=0.91$), but reached this level faster with \textbf{Claude} (v7 vs.\ v5). This suggests same-family advantages may manifest as faster iteration rather than higher ceilings.

\textbf{Codex maintains peak performance across classifiers.} While \textbf{Codex} required more iterations with \textbf{Claude} classifier, it ultimately achieved the same $\kappa=0.93$. This suggests \textbf{Codex}'s iterative refinement strategy is robust across model families, though less efficient when the classifier differs from its native architecture.

\textbf{Practical implications.} When deploying LLM-based labeling, researchers should consider potential interactions between the iterating agent and the classifier model. Cross-classifier validation (testing prompts developed with one classifier on another) may reveal brittleness in prompt engineering solutions.

\section{Experiment 4: Gemini Classifier}

\subsection{Motivation}

Experiment 3 showed that classifier choice affects performance differently for each agent. With only two classifiers, we could not fully characterize these patterns. To complete the picture, we conducted a third replication using \textbf{Gemini 3 Pro} as the classifier.

\subsection{Method}

We replicated the cross-agent validation using \textbf{Gemini 3 Pro} Preview as the underlying classifier. The same three agents (\textbf{Codex}, \textbf{Claude Code}, \textbf{Gemini}) independently iterated for 11 versions (v0--v10) with identical instructions. This allowed us to test whether \textbf{Gemini Agent} would benefit from its same-family classifier.

\subsection{Results}

\begin{table}[t]
\centering
\renewcommand{\arraystretch}{1.2}
\caption{Best performance across all three classifiers.}
\label{tab:three_classifiers}
\begin{tabular}{lllll} \hline
Agent & \textbf{GPT} & \textbf{Claude} & \textbf{Gemini} & Same-Family Best? \\ \hline
\textbf{Codex} (OpenAI) & \textbf{0.93} & \textbf{0.93} & 0.87 & $\checkmark$ \textbf{GPT} \\
\textbf{Claude Code} & \textbf{0.91} & \textbf{0.91} & 0.89 & $\checkmark$ \textbf{GPT}/\textbf{Claude} \\
\textbf{Gemini Agent} & \textbf{0.93} & 0.85 & 0.92 & $\checkmark$ \textbf{GPT}/\textbf{Gemini} \\ \hline
\end{tabular}
\end{table}

Table~\ref{tab:three_classifiers} shows the complete comparison across all three classifiers. \textbf{Gemini Agent} achieved strong performance with both \textbf{GPT} classifier ($\kappa=0.93$) and its same-family \textbf{Gemini} classifier ($\kappa=0.92$), demonstrating that iterative refinement can overcome classifier family differences. \textbf{Codex} showed clear same-family preference, dropping from $\kappa=0.93$ with \textbf{GPT} to $0.87$ with \textbf{Gemini} classifier. \textbf{Claude Code} achieved identical performance with \textbf{GPT} and \textbf{Claude} classifiers ($\kappa=0.91$), with slightly lower performance on \textbf{Gemini} classifier ($\kappa=0.89$).

\begin{figure*}[t]
\Description{Three-panel figure showing classifier comparison. Left panel shows bar chart of best performance by classifier. \textbf{Gemini Agent} achieves 0.93 with \textbf{GPT} and 0.92 with \textbf{Gemini} classifier. \textbf{Codex} drops from 0.93 to 0.87 when switching from \textbf{GPT} to \textbf{Gemini} classifier, showing same-family preference.}
\centering
\includegraphics[width=0.95\textwidth]{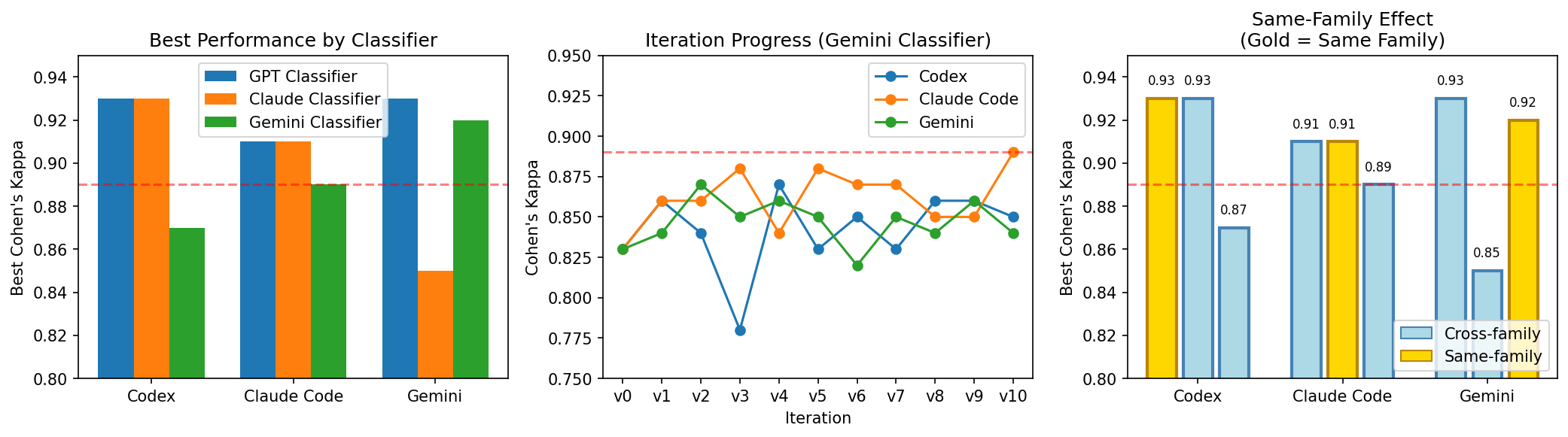}
\caption{Three-classifier comparison. \textbf{Gemini Agent} achieves strong performance across classifiers ($0.93$ \textbf{GPT}, $0.92$ \textbf{Gemini}). \textbf{Codex} shows clear same-family preference ($0.93$ \textbf{GPT} vs.\ $0.87$ \textbf{Gemini}). \textbf{Claude Code} achieves consistent performance across \textbf{GPT} and \textbf{Claude} classifiers ($0.91$).}
\label{fig:three_classifiers}
\end{figure*}

Figure~\ref{fig:three_classifiers} visualizes the iteration dynamics and same-family effects across all three classifiers.

\subsection{Interpretation}

The three-classifier comparison reveals nuanced patterns:

\textbf{Same-family advantages manifest as faster convergence.} \textbf{Gemini Agent} reached $\kappa=0.92$ at v5 with \textbf{Gemini} classifier, requiring only v9 for the higher $0.93$ with \textbf{GPT}. Similarly, \textbf{Claude Code} achieved $0.91$ at v7 with \textbf{Claude} classifier (Experiment 3), comparable to v5 for $0.91$ with \textbf{GPT}. \textbf{Codex} showed the clearest same-family effect: stable convergence with \textbf{GPT} but volatile iteration patterns with non-\textbf{GPT} classifiers.

\textbf{GPT classifier enables highest performance.} Both \textbf{Codex} and \textbf{Gemini Agent} achieved their peak performance ($\kappa=0.93$) with \textbf{GPT} classifier, suggesting \textbf{GPT-5.2} may be particularly well-suited for this labeling task or more responsive to iterative prompt refinement.

\textbf{No universal ``best'' classifier.} Each agent-classifier combination produces different results. This argues for testing multiple classifiers when developing LLM-based labeling pipelines, rather than assuming any single classifier is universally optimal.

\section{Error Analysis}

Analyzing classifier disagreements with human consensus revealed four primary error categories. The largest category, \textbf{Ambiguous Intent} (38\%), comprised sessions that genuinely straddled category boundaries; notably, human coders also disagreed on these cases, suggesting inherent ambiguity rather than classifier failure. \textbf{Context-Dependent Interpretation} (27\%) involved phrases whose meaning shifted based on surrounding context, such as ``I don't understand'' signaling either confusion or frustration depending on prior exchanges. \textbf{Multi-Turn Dynamics} (22\%) captured sessions where student intent evolved mid-conversation, challenging session-level labels. Finally, \textbf{Rare Patterns} (13\%) represented unusual interactions appearing only once or twice in the dataset. These error patterns suggest productive future directions: multi-label approaches for genuinely ambiguous sessions, explicit conversation history modeling for context-dependent interpretation, and turn-level rather than session-level labeling for dynamic interactions.

\section{Discussion}

\subsection{From Tool to Teammate: A Shift in Workflow}

Traditional approaches treat LLMs as tools to be configured; our approach treats coding agents as \textit{teammates} that analyze errors, propose fixes, and document reasoning. This shift in workflow enables qualitatively different practices. At its core, the technical mechanism is iterative prompt refinement with validation feedback, related to automated prompt optimization methods such as APE~\cite{zhou2023large}, DSPy~\cite{khattab2023dspy}, and OPRO~\cite{yang2023large}. The key distinction is that coding agents perform qualitative error analysis and propose theory-grounded modifications with documented reasoning, enabling researcher oversight at each step rather than optimizing a metric through opaque search. Agent-driven iteration reached 20+ versions compared to the typical 3--5 manual iterations, enabling more thorough exploration of the solution space. Agents systematically categorize all disagreements without the selective attention that can bias human analysis. Transparent reasoning throughout the process enables reproducibility in ways that ``we tuned the prompt'' cannot. Most valuably, the ``confusion = engagement'' insight emerged from human-AI disagreement, demonstrating how this collaborative process can surface tacit pedagogical knowledge that researchers hold implicitly but have not articulated.

\subsection{Agent Performance and Classifier Effects}

A notable finding from Experiment 2 is that all three agents achieved strong reliability with \textbf{GPT-5.2} as classifier: \textbf{Codex} and \textbf{Gemini} both achieved $\kappa=0.93$, while \textbf{Claude Code} reached $\kappa=0.91$. Experiments 3--4 revealed how classifier choice affects each agent differently. Across all three classifiers, each agent showed distinct patterns: \textbf{Codex} performed best with \textbf{GPT} and \textbf{Claude} classifiers (both $\kappa=0.93$) but dropped to $0.87$ with \textbf{Gemini}; \textbf{Claude Code} achieved consistent performance with \textbf{GPT} and \textbf{Claude} classifiers ($\kappa=0.91$), with slightly lower performance on \textbf{Gemini} ($\kappa=0.89$); \textbf{Gemini Agent} performed best with \textbf{GPT} classifier ($0.93$), with strong performance also on \textbf{Gemini} ($0.92$) but lower on \textbf{Claude} ($0.85$).

The ``same-family advantage'' manifests primarily as \textit{faster convergence} rather than higher ceiling performance. \textbf{Claude Code} reached $\kappa=0.91$ at v7 with \textbf{Claude} classifier, similar to v5 for $0.91$ with \textbf{GPT}. \textbf{Gemini Agent} achieved $\kappa=0.92$ at v5 with \textbf{Gemini} classifier, versus v9 for $0.93$ with \textbf{GPT}. \textbf{Codex} showed the most dramatic pattern: smooth convergence to $0.93$ at v7 with \textbf{GPT}, but more volatile iteration with \textbf{Claude} classifier before reaching $0.93$ at v10. These results suggest same-family advantages enable more efficient and stable iteration, even when cross-family classifiers ultimately achieve comparable or higher performance. In practical terms, researchers facing time constraints may benefit from same-family pairings that converge in fewer iterations, while those prioritizing maximum performance can invest additional iterations with cross-family classifiers.

\subsection{Prompt Iteration Strategies}

Analyzing the prompts developed by each agent reveals distinct strategies. \textbf{Codex} developed explicit, ordered decision rules with clear priority (e.g., ``S only if clearly unrelated topic... E otherwise''). \textbf{Claude Code} independently converged on nearly identical rules, supporting the hypothesis that systematic iteration discovers genuine patterns rather than agent-specific artifacts. \textbf{Gemini} used heuristic guidance (``Consider the context. If the student seems frustrated, it's more likely EA'') instead of strict rules, leaving more interpretation to the classifier.

The correlation between strategy and performance is notable: \textbf{Codex}'s explicit rules achieved $\kappa=0.83$ on Follow-up Type, while \textbf{Gemini}'s heuristics plateaued at $\kappa=0.53$. For challenging dimensions, explicit disambiguation rules outperform intuitive guidelines. All prompts and iteration logs are available in supplementary materials.

\subsection{Iteration Dynamics: Two Regression Patterns}

Our experiments reveal two distinct regression patterns illustrating why per-iteration validation is essential.

\textbf{Pattern 1: Over-correction.} With the \textbf{Gemini} classifier, \textbf{Codex} experienced a sharp drop at v3 (from $\kappa=0.85$ to $\kappa=0.78$). Analysis revealed the cause: v3 restructured decision priority from ``AS $>$ HL $>$ OT'' to ``OT $>$ AS $>$ HL,'' adding extensive jailbreak detection rules. This over-emphasis caused the classifier to misclassify legitimate help-seeking behaviors as ``Other.'' The agent recognized the regression at v4 and partially reverted, recovering to $\kappa=0.85$.

\textbf{Pattern 2: Output format instability.} With the \textbf{Claude} classifier, \textbf{Codex} dropped from $\kappa=0.86$ to $\kappa=0.60$ at v7. The cause differed: the agent aggressively shortened the prompt to reduce token costs, destabilizing output format. Only 7 of 80 sessions were successfully parsed at v7, compared to 39 at v6. The agent detected this through validation metrics and expanded critical sections at v8, recovering to $\kappa=0.84$.

These patterns highlight complementary failure modes: over-correction alters classification behavior in unintended ways; output instability affects structured output reliability. Both underscore the importance of per-iteration validation.

\subsection{When to Stop Iterating}

These regression patterns raise the question: when should researchers stop iterating? Several indicators signal diminishing returns. Performance plateaus are the most obvious: if two consecutive iterations show $\Delta\kappa < 0.02$, additional refinement is unlikely to help. Human inter-rater reliability provides a natural ceiling. Error analysis offers another diagnostic: when remaining disagreements cluster in genuinely ambiguous cases, the prompt has likely extracted most learnable signal. Prompt iteration should follow standard machine learning practice: use a held-out validation set to select the best-performing version, rather than assuming the latest iteration is always superior.

\subsection{Practical Recommendations}

Our experience suggests several principles for researchers adopting this methodology. A well-developed codebook provides the strongest foundation, as clear category definitions transfer remarkably well to LLM classifiers. Starting with minimal prompts and adding complexity only when specific errors demand it prevents premature overengineering. Researchers should expect 5--10 iterations for most tasks, focusing on the weakest dimension rather than improving everything simultaneously. Documenting failures proves valuable, as they often reveal unstated assumptions. For validation, reserving held-out test sets catches overfit prompts, and human inter-rater reliability serves as the appropriate benchmark. Given the effectiveness of iterative refinement demonstrated in our experiments, we recommend that future prompt engineering research adopt similar validation practices, with explicit documentation of prompt versions, per-iteration metrics, and held-out test results.

\subsection{Limitations and Generalizability}

The labeling framework was designed for mathematical tutoring dialogues; while the core methodology (agent-as-teammate workflow, iteration stopping criteria, cross-validation approach) transfers to other domains, specific prompt rules would require adaptation. The procedural/conceptual distinction in Topic Type, for instance, reflects mathematical problem-solving; writing tutoring might instead distinguish revision feedback from content guidance.

The 80-session validation set limits statistical power for detecting subtle differences and creates potential for overfitting, as agents iterated on the same set used for evaluation. Future work should implement strict train/validation/test splits. Our classifier comparisons examined only three model families; other combinations may yield different patterns as models evolve.

Our classification operates at the session level, assigning a single label per dimension to each tutoring session. This approach does not capture within-session dynamics where student intent may shift (e.g., from help-seeking to answer-seeking mid-conversation). Turn-level labeling with sequential modeling would enable finer-grained analysis and real-time adaptive interventions, and represents an important direction for future work.

We did not compare against fine-tuned models, which may achieve higher performance on this task given sufficient training data. However, fine-tuning requires substantially more labeled examples than our 80-session set and lacks the interpretability benefits of explicit prompt rules that can be reviewed and validated by domain experts. Future work should compare iterative prompt refinement against fine-tuning approaches as labeled datasets grow.

While we used temperature$=0$ for all classifier calls to minimize stochastic variation, LLM outputs are not fully deterministic across API calls. Current commercial LLM APIs do not support seed-based reproducibility for chat completions. We mitigate this limitation through cross-validation across multiple folds, which provides robustness against single-run artifacts.

Our evaluation lacks a fully independent final test set separated from all tuning phases. While 4-fold cross-validation provides held-out estimates, the best-performing prompt version was selected based on development-set performance rather than evaluated on a completely untouched partition. This means our reported cross-validated $\kappa=0.78$ represents generalization within the 80-session pool, not performance on entirely unseen data. Future work should reserve a separate test set from the broader 659-session corpus for unbiased final evaluation.

The 80-session validation set also limits the statistical reliability of performance comparisons between agents and classifiers. With fold sizes of 20 sessions and standard deviations of $0.08$--$0.12$ across folds, observed differences between agents (e.g., $\kappa=0.78$ vs.\ $0.76$) may not be statistically distinguishable. We report descriptive comparisons rather than formal hypothesis tests, and caution against interpreting small $\kappa$ differences as meaningful. Larger validation sets would enable confidence intervals and significance testing to determine whether agent or classifier differences are reliable.

\subsection{Implications for Educational Research and Practice}

Our methodology enables educational dialogue analysis at scale. Traditional behavioral coding is limited to tens or hundreds of sessions; with LLM agent collaboration, researchers can label thousands of sessions at modest cost (\$5--8 per coding scheme), enabling longitudinal and cross-institutional studies previously infeasible.

Our findings demonstrate \textit{appropriate reliance}~\cite{vasconcelos2023explanations}: humans provided theoretical grounding while AI provided systematic iteration. The ``confusion = engagement'' discovery illustrates how AI can surface tacit knowledge, suggesting LLM agents can serve as collaborators in codebook development. For tutoring system design, real-time detection of answer-seeking could trigger adaptive interventions~\cite{kapur2008productive}.

\subsection{LLMs as Inter-Rater Reliability Partners}

Our results suggest LLMs as collaborative partners for inter-rater reliability. Traditional IRR requires multiple trained human coders; our experiments show LLM classifiers achieving $\kappa=0.91$--$0.93$, enabling hybrid workflows where a single human partners with LLM classifiers. This could make rigorous qualitative coding accessible to teams lacking resources for traditional multi-coder approaches.

\section{Conclusion}

We presented a methodology where LLM coding agents serve as research teammates for behavioral labeling in educational dialogue analysis. Our primary finding is that 4-fold cross-validation on held-out data confirmed genuine improvement: the best agent achieved test $\kappa=0.78$ (SD$=0.08$), matching human inter-rater reliability ($\kappa=0.78$). While development-set performance reached $\kappa=0.91$--$0.93$, these cross-validated results represent the generalization claim. In Experiment 1, a single agent improved Follow-up Type from $\kappa=0.59$ to $\kappa=0.68$ over 21 iterations. In Experiment 2, three independent agents all improved from baseline, with \textbf{Codex} and \textbf{Gemini} reaching $\kappa=0.93$ on the development set. Experiments 3--4 revealed that classifier choice matters: same-family classifiers enabled faster convergence, while \textbf{GPT} classifier achieved the highest ceiling performance. Continued iteration beyond the optimum led to regression, highlighting the importance of held-out validation. Throughout, costs remained modest at approximately \$5--8 per agent per classifier, with total time under one week of part-time researcher effort. Systematic disagreement analysis also surfaced implicit labeling criteria that human coders applied but had not documented, demonstrating how this collaborative process can uncover tacit expert knowledge.

Treating LLM agents as collaborative partners rather than static tools enables qualitatively different research workflows. Twenty-plus iterations become routine rather than exceptional, and process transparency becomes the norm rather than the exception. However, as with any optimization process, researchers must use held-out validation sets to identify the best-performing prompt, following standard machine learning practice.

\section*{Code Availability}

Iteration logs, prompts, and evaluation scripts are available at: \url{https://github.com/EasonC13/agent-prompt-iteration}

\bibliographystyle{abbrv}
\bibliography{merged-references}

\end{document}